\documentclass[conference]{IEEEtran}

\usepackage{cite}

\ifCLASSINFOpdf
  \usepackage[pdftex]{graphicx}
\else
\fi
\graphicspath{{figures/}}

\usepackage[cmex10]{amsmath}
\usepackage{amssymb}
\usepackage{mathtools}
\DeclareMathOperator{\Tr}{{tr}}

\DeclareMathOperator{\rank}{{rank}}
\usepackage{url}

\usepackage{epstopdf}					
\usepackage{amsthm}

\theoremstyle{plain}

\usepackage{bm}

\usepackage{xcolor}

\begin{document}
%
\title{\LARGE{Low-Complexity Detection of $M$-ary PSK Faster-than-Nyquist Signaling}}

\author{\IEEEauthorblockN{Ebrahim Bedeer}
\IEEEauthorblockA{\textit{School of Engineering} \\
\textit{Ulster University}\\
Newtownabbey, United Kingdom \\
e.bedeer.mohamed@ulster.ac.uk}
\and
\IEEEauthorblockN{Halim Yanikomeroglu}
\IEEEauthorblockA{\textit{System and Computer Engineering Dept.} \\
\textit{Carleton University}\\
Ottawa, ON, Canada \\
halim.yanikomeroglu@sce.carleton.ca}
\and
\IEEEauthorblockN{Mohamed Hossam Ahmed}
\IEEEauthorblockA{\textit{Faculty of 	Engineering and Applied Science} \\
\textit{Memorial University}\\
St. John's, NL, Canada \\
mhahmed@mun.ca}}

\maketitle

\begin{abstract}
Faster-than-Nyquist (FTN) signaling is a promising non-orthogonal physical layer transmission technique to improve the spectral efficiency of future communication systems but at the expense of intersymbol-interference (ISI). 
In this paper, we investigate the detection problem of FTN signaling and formulate the sequence estimation problem of any $M$-ary phase shift keying (PSK) FTN signaling as an optimization problem that turns out to be non-convex and nondeterministic polynomial time (NP)-hard to solve. We propose a novel algorithm, based on concepts from semidefinite relaxation (SDR) and Gaussian randomization, to detect any $M$-ary PSK FTN signaling in polynomial time complexity regardless of the constellation size $M$ or the ISI length. Simulation results show that the proposed algorithm strikes a balance between the achieved performance and the computational complexity. Additionally, results show the merits of the proposed algorithm in improving the spectral efficiency when compared to Nyquist signaling and the state-of-the-art schemes from the literature. In particular, when compared to Nyquist signaling at the same error rate and signal-to-noise ratio, our scheme provides around $17\%$ increase in the spectral efficiency at a roll-off factor of 0.3.
\end{abstract} 



\section{Introduction}
In the last decades, we witnessed an exponential growth of wireless traffic that is expected to continue increasing in future communication systems. That said, improving the spectral efficiency (SE) becomes crucial given the scarcity of the available spectrum. 
Faster-than-Nyquist (FTN) signaling \cite{anderson2013faster} is one of the candidate solutions to increase the SE of future communication networks. In FTN signaling, the pulses' transmission rate exceeds the Nyquist limit, and this results in intersymbol-interference (ISI) between the received pulses.
James Mazo \cite{mazo1975faster} proved that the minimum distance between the received pulses will not be reduced if the signaling rate, that exceeds the Nyquist limit, is within what is known as Mazo limit (sinc pulses can be packed up to 0.802 of its symbol duration).
There were doubts raised by Foschini about the benefits of FTN signal in \cite{foschini1984contrasting}; however, it potentials were revealed in \cite{rusek2006cth04, rusek2009constrained} when Rusek and Anderson showed that lower bounds on the information rate of FTN signalling are in most cases higher than the information rates of conventional Nyquist signalling. Thenceforth, the concept of FTN signalling has been extended in many directions: different pulse shapes \cite{liveris2003exploiting}, frequency-domain \cite{rusek2005two, rusek2009multistream}, non-binary signalling \cite{wang1995practically, rusek2008non}, non-orthogonal pulses \cite{seshadri1988asymptotic}, and multiple-input-multiple-output channels \cite{rusek2009existence}. 


Low-complexity detection of FTN signaling attracted the attention of research community in recent years. For instance, the authors in  \cite{bedeer2016reduced} proposed a technique to reduce the search space of the maximum likelihood sequence estimation (MLSE) algorithm, and hence, attain the optimal performance at reduced complexity. 
Suboptimal solutions of the MLSE detection problem were reported in \cite{anderson2009new, prlja2012reduced}, where reduced state or reduced search  trellis coding-based Bahl-Cocke-Jelinek-Raviv (BCJR) algorithms were proposed, respectively.
The computational complexity of the works in \cite{bedeer2016reduced, prlja2008receivers, anderson2009new, prlja2012reduced} are still exponential with the ISI length, and extending them to high-order modulation will be problematic. In \cite{bedeer2017low}, the authors proposed a polynomial-time complexity sub-optimal detection algorithm of any high-order quadrature amplitude modulation (QAM) FTN signaling. A novel FTN detector for binary phase shift keying (BPSK) and quadrature PSK (QPSK) on a symbol-by-symbol basis is proposed in \cite{bedeer2017very}, where the computational complexity is independent of the ISI length.

{{$M$-ary phase shift keying (PSK) modulates only the phase of symbols, and hence, the transmitter can always operate at peak power (unlike $M$-ary QAM). This is why $M$-ary PSK is preferred in constant amplitude transmission for satellite communications \cite{corazza2007digital}.  The benefits of employing FTN for satellite communications were recently explored in \cite{beidas2014faster}.
That said,  this paper investigates the detection problem of $M$-ary phase shift keying (PSK) FTN signaling.}} We formulate the MLSE problem of any $M$-ary PSK FTN signaling as an optimization problem that turns out to be non-convex and nondeterministic polynomial time (NP)-hard to solve. We used concept of semidefinite relaxation to produce a number of relaxed convex optimization problems, where the global optimal solution can be reached in polynomial time complexity regardless of the constellation size $M$ or the ISI length. Then, with the help of Gaussian randomization, we obtain a feasible solution to the original MLSE of $M$-ary PSK FTN signaling problem from the relaxed convex problem solution. We present simulation results to validate our theoretical analysis.

The paper is organized as follows.
The $M$-ary PSK FTN signaling system model and the detection problem are discussed in Section~\ref{sec:model}; while the proposed SDR technique is investigated in Section~\ref{sec:SDP}. The numerical results are presented in Section~\ref{sec:results} and finally the paper is concluded in Section~\ref{sec:conc}.


\section{System Model and FTN Signaling Detection Problem Formulation} \label{sec:model} 

\begin{figure*}[!t]
	\centering
	\includegraphics[width=0.750\textwidth]{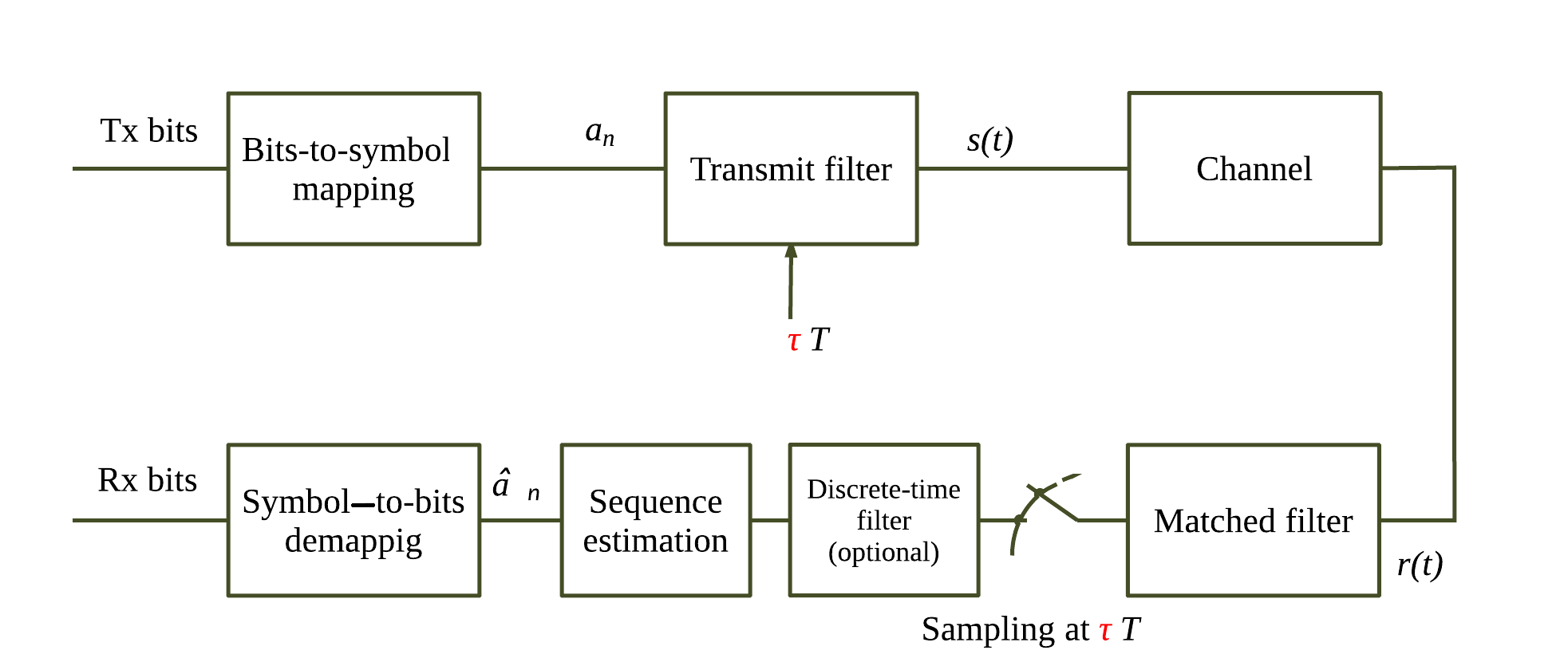}
	\caption{Block diagram of FTN signaling.}\label{fig:block_diagram}
\end{figure*}

Figure \ref{fig:block_diagram} shows the block diagram of a  baseband communication system employing FTN signaling. At the transmitter, the data symbols are drawn from a $M$-ary PSK constellation and then shaped by a unit-energy pulse $p(t)$. We transmit a total of $N$ symbols, each every $\tau T$, where $\tau \in (0,1]$ is the time packing factor and $T$ is the symbol duration.
The received signal -- after being affected by additive white Gaussian noise (AWGN) channel -- is passed through a filter matched to the transmit pulse, and is given as

\begin{eqnarray}
{y}(t) = \sqrt{\tau \: E_s} \: \sum\nolimits_{n = 1}^{N} {a}_n g(t - n \tau T) + {q}(t),
\end{eqnarray}
where ${a}_n, \: n= 1, \hdots, N,$ is the complex $M$-ary PSK data symbol, $E_s$ is the data symbol  energy, $g(t) = \int\nolimits p(x) p(x - t) dx$, ${q}(t) = \int\nolimits n(x) p(x - t) dx$ with $n(t)$ is the AWGN with zero mean and variance $\sigma^2$, and $1/(\tau T)$ is the signaling rate. 
The output signal of the matched-filter is sampled every $\tau T$ and can be expressed as 
\begin{eqnarray}\label{eq:ISI_initial}
{y}_k 
&=& \sqrt{\tau \:E_s} \sum\nolimits_{n = 1}^{N} {a}_n g(k \tau T - n \tau T) + {q}(k \tau T), \nonumber \\
&=& \underbrace{\sqrt{\tau \:E_s} \: {a}_k \: g(0)}_{\textup{desired symbol}} \nonumber \\ & & + \underbrace{\sqrt{\tau \: E_s} \: \sum\nolimits_{n = 1, \: n \ne k}^{N} {a}_n \: g((k - n) \tau T)}_{\textup{ISI}}  + \: {q}(k \tau T). \nonumber \\
\end{eqnarray} 
{
We rewrite  \eqref{eq:ISI_initial} in a vector form as 
\begin{eqnarray}\label{eq:vector_RX}
{\bm{y}}_{\textup{c}} &=& \sqrt{\tau \: E_s} \: {\bm{a}} \star \bm{g} + {\bm{q}}_{\textup{c}}, 
\end{eqnarray} 
{{where $\bm{a}$, $\bm{g}$, $\bm{q}_c$, and $\star$ are the transmit data symbol vector, ISI vector, colored noise vector, and the convolution operator, respectively.}}
One can see from \eqref{eq:ISI_initial} that the noise samples are correlated, and in this paper, we follow the approach of designing a whitening matched filter at the receiver of the FTN signaling \cite{prlja2008receivers}.
In particular, we use spectral factorization in order to find the approximate whitening filter coefficients $1/V(1/z^*)$ from the vector $\bm{g}$. This is achieved by  factorizing $G(z)$ into $V(z) V(1/z^*)$ \cite{oppenheim2010discrete}, where $G(z)$ is the z-transform of  $\bm{g}$. {For more details on the design of the whitened matched filter for FTN receivers, we refer the reader to  \cite{bedeer2017low, prlja2008receivers}.}
Accordingly, the received signal in \eqref{eq:vector_RX}, after passing through the approximate whitening filter $1/V(1/z^*)$, is given by 
\begin{eqnarray}\label{eq:complex}
{\bm{y}}_{\textup{w}} &=& \sqrt{\tau \: E_s} \: {\bm{a}} \star \bm{v} + {\bm{q}}_{\textup{w}}, 
\end{eqnarray}  
where ${\bm{q}}_{\textup{w}}$ is the white Gaussian noise $\sim \mathcal{N}(0,\sigma^2)$ 
and the causal ISI $\bm{v}$ is constructed from $\bm{g}$ as  $\bm{v}[n] \star \bm{v}[-n] = \bm{g}$. 
Equation \eqref{eq:complex} is equivalently expressed as
\begin{eqnarray}\label{eq:complex_model_white}
{\bm{y}}_{\textup{w}} &=&  {\bm{V}} {\bm{a}}  + {\bm{q}}_{\textup{w}}, 
\end{eqnarray}  
where ${\bm{V}}$ is the ISI matrix constructed from the vector $\bm{v}$. Finally, the MLSE problem to detect $M$-ary PSK FTN signaling  is written as
\begin{eqnarray}\label{eq:opML_white}
\underset{\bm{a}  \: \in \: \mathcal{D}}{\min}  & &   \Vert \bm{y}_{\textup{w}} - \bm{V} \bm{a} \Vert^2_2 \nonumber \\
{\textup{s. t.}} & &  \mathcal{D} = \left\{a = e^{j\frac{(2m - 1)\pi}{M}} | m = 1, ..., M\right\},
\end{eqnarray} 
where $\Vert . \Vert_2$ is the $2$-norm. The constraint in \eqref{eq:opML_white} denotes that the decision variables are discrete as they are drawn from an $M$-ary PSK constellation.
Exhaustive search can be used to solve the MLSE problem in \eqref{eq:opML_white}; however, it is too complex to be implemented in practical systems.
In the following section, we propose a suboptimal reduced-complexity detection algorithm to detect $M$-ary PSK FTN signaling based on concepts from semidefinite relaxation and Gaussian randomization.
}
\section{Proposed Algorithm to detect $M$-ary PSK FTN Signaling} \label{sec:SDP}

We propose our semidefinite relaxation based algorithm for the detection problem of any $M$-ary PSK FTN signaling in this section. As it will be discussed, the proposed scheme provides a sub-optimal solution at polynomial time complexity regardless of the constellation size $M$ or the ISI length.

The objective of the   problem in~\eqref{eq:opML_white} can be re-written as
\begin{eqnarray}\label{eq:obj}
\Vert \bm{y}_{\textup{w}} \Vert_2^2 - 2 \Re\{\bm{a}^{\textup{H}} \bm{V}^{\textup{H}} \bm{y}_{\textup{w}}\} + \Tr\{\bm{V}^{\textup{H}} \bm{V} \bm{A}\}, 
\end{eqnarray}
where $\Re\{.\}$ represents the real part of complex numbers and $\bm{A} = \bm{a} \bm{a}^{\textup{H}}$, where $(.)^{\textup{H}}$ denotes the conjugate transpose.  Accordingly, the MLSE problem in \eqref{eq:opML_white} to detect $M$-ary PSK FTN signaling can be re-formulated as
\begin{eqnarray}\label{eq:SDP}
\underset{\bm{a}  \: \in \: \mathcal{D}, \: \bm{A} \in \mathbb{C}^N}{\min}  & &   \Tr\{\bm{V}^{\textup{H}} \bm{V} \bm{A}\} - 2 \Re\{\bm{a}^{\textup{H}} \bm{V}^{\textup{H}} \bm{y}_{\textup{w}}\} \nonumber \\
{\textup{s. t. \hfill}} & &  \mathcal{D} = \left\{a = e^{j\frac{(2m - 1)\pi}{M}} | m = 1, ..., M\right\}, \nonumber \\
& & \bm{A} = \bm{a} \bm{a}^{\textup{H}},
\end{eqnarray} 
where $\mathbb{C}^N$ denotes the set of all complex $N \times N$ Hermitian matrices.  Please note that the term $\Vert \bm{y}_{\textup{w}} \Vert_2^2$ in \eqref{eq:obj} is dropped from the objective function in \eqref{eq:SDP} as it is not a function of the decision variables, and hence, is  constant. 
Please note that we define a new variable $\bm{A} = \bm{a} \bm{a}^{\textup{H}}$, which is equivalent to have the matrix $\bm{A}$ being a rank one ($\rank\{\bm{A}\} = 1$) Hermitian positive semidefinite ($\bm{A} \succeq 0$).
The problem in \eqref{eq:SDP} is still NP-hard and the main difficulties lie in the non-convex rank one and the discrete constellation constraints. In the following, we investigate and propose techniques to properly address these non-convex constraints.

The objective function in \eqref{eq:SDP} is linear in the matrix $\bm{A}$ and data symbols vector $\bm{a}$. However, and as discussed earlier, the reformulation from \eqref{eq:opML_white} to \eqref{eq:SDP} allows us to reveal one of the  difficulties in solving the MLSE optimization problem to detect the $\bm{M}$-ary PSK FTN signaling. It is clear from \eqref{eq:SDP} that the difficulty lies in the nonconvex rank one constraint. We relax this problem into a convex one by replacing the nonconvex rank one constraint with a convex positive semidefinite constraint, i.e., $\bm{A} - \bm{a} \bm{a}^{\textup{H}} \succeq 0$ \cite{luo2010semidefinite}. 
Further, we substitute this positive semidefinite constraint, i.e., $\bm{A} - \bm{a} \bm{a}^{\textup{H}} \succeq 0$, with its schur complement \cite{Boyd2004convex}. Hence, the semidefinite relaxation problem of detecting  $M$-ary FTN signaling is written as
\begin{eqnarray}\label{eq:SDR_2}
\underset{\bm{a}  \: \in \: \mathcal{D}, \: \bm{A} \in \mathbb{C}^N}{\min}  & &   \Tr\{\bm{V}^{\textup{H}} \bm{V} \bm{A}\} - 2 \Re\{\bm{a}^{\textup{H}} \bm{V}^{\textup{H}} \bm{y}_{\textup{w}}\} \nonumber \\
{\textup{s. t.}} & &  \mathcal{D} = \left\{a = e^{j\frac{(2m - 1)\pi}{M}} | m = 1, ..., M\right\}, \nonumber \\
& & \begin{bmatrix}
\bm{A} & \bm{a}\\ 
\bm{a}^{\textup{H}} & 1
\end{bmatrix} \succeq 0.
\end{eqnarray} 
The difficulty now in solving \eqref{eq:SDR_2} lies in the discrete constellation constraint. To tackle such difficulty, we follow a similar approach to the one presented in \cite{so2007approximating} and relax the discrete constellation constraint (with discrete decision variables) to continuous ones that lie on a unit circle. Such relaxation is guaranteed to achieve a sub-optimal solution  within a certain accuracy of the solution of the original MLSE problem, at reduced computational complexity. Please note that quantifying the gap to the optimal solution is outside the scope of the this work; however, we refer the interested reader to \cite{so2007approximating} for additional discussion on how to calculate such gap. That said, the problem in \eqref{eq:SDR_2} is rewritten as
\begin{eqnarray}\label{eq:SDR_3}
\underset{\bm{a}  \: \in \: \mathcal{D}, \: \bm{A} \in \mathbb{C}^N}{\min}  & &   \Tr\{\bm{V}^{\textup{H}} \bm{V} \bm{A}\} - 2 \Re\{\bm{a}^{\textup{H}} \bm{V}^{\textup{H}} \bm{y}_{\textup{w}}\} \nonumber \\
{\textup{s. t. \hfill}} & & \vert a_n \vert = 1, \qquad n = 1, \hdots, N,  \nonumber \\
& & \begin{bmatrix}
\bm{A} & \bm{a}\\ 
\bm{a}^{\textup{H}} & 1
\end{bmatrix} \succeq 0.
\end{eqnarray} 
It is shown in \cite{grant2010cvx} that the convex problem in \eqref{eq:SDR_3} can be solved efficiently using numerical solvers. 
The challenging task  is to convert the global solutions $\bm{A}^*$ and $\bm{a}^*$ of the convex SDR problem in \eqref{eq:SDR_3} into a feasible solution $\hat{\bm{a}}$ to the original $M$-ary PSK FTN signaling in \eqref{eq:SDP}. 
Among several methods that can achieve this task, we use a Gaussian randomization method inspired by the works in \cite{so2007approximating, luo2010semidefinite} due to its efficiency and low complexity. It is worthy to emphasize that Gaussian randomization, as well as other methods used to obtain feasible solution to the original detection problem that is NP-hard, may not achieve the global optimal solution. This is not surprising as we would have solved an NP-hard problem in a polynomial time complexity.
Gaussian randomization solves a stochastic version of \eqref{eq:SDR_3}, and the sub-optimal solution $\hat{\bm{a}}$ to \eqref{eq:SDP} is selected to be the solution that minimizes the objective in \eqref{eq:SDR_3} among candidate solutions generated from a multivariate Gaussian random vectors with mean $\bm{a}^*$ and covariance matrix ($\bm{A}^* - \bm{a}^* \bm{a}^{*\textup{H}}$).
The Gaussian randomization procedure can be explained as follows. {{We generate ${\bm{\xi}}_{\ell}, \ell = 1, ..., L,$ where $L$ is the total number of randomization iterations, as ${\bm{\xi}}_{\ell} \sim \mathcal{N}(\bm{a}^*, \bm{A}^* - \bm{a}^* \bm{a}^{*,\textup{H}})$, then we find $\breve{\bm{a}}_{\ell}$ as follows}} \cite{so2007approximating}:
\begin{IEEEeqnarray}{rcl}\label{eq:subopt}
\breve{{a}}_{\ell} = \hspace{8cm}\nonumber \\ \left\{\begin{matrix}
1 & \textup{if} \arg({{\xi}}_{\ell}) \in [-\pi/m, \pi/m) \\ 
e^{j 2\pi/M} & \textup{if} \arg({{\xi}}_{\ell}) \in [\pi/m, 3\pi/m)\\ 
\vdots & \vdots\\ 
e^{j 2\pi(M-1))/M} &  \textup{if} \arg({{\xi}}_{\ell}) \in [(2m - 3)\pi/m, (2m - 1)\pi/m).
\end{matrix}\right. \nonumber \\
\end{IEEEeqnarray}
The optimal value of $\ell$, i.e., $\ell_{\textup{op}}$, is selected to minimize the objective function in \eqref{eq:SDR_3} as
\begin{IEEEeqnarray}{rcl}\label{eq:temp}
\ell_{\textup{op}} & = & \arg \underset{\ell = 1, ..., L}{\min} \Tr\{\bm{V}^{\textup{H}} \bm{V} \breve{\bm{a}}_{\ell} \breve{\bm{a}}_{\ell}^{\textup{H}}\} - 2 \Re\{\breve{\bm{a}}_{\ell}^{\textup{H}} \bm{V}^{\textup{H}} \bm{y}_{\textup{w}}\}, 
\end{IEEEeqnarray}
and then, the sub-optimal solution vector $\hat{\bm{a}}$ is identified as 
\begin{eqnarray}
\hat{\bm{a}} & = & \breve{\bm{a}}_{\ell_{\textup{op}}}.
\end{eqnarray}

The proposed algorithm\footnote{Please note that the proposed algorithm can be extended to produce soft-outputs, and hence, to be used with channel coding. This can be achieved by exploiting Gaussian randomization to efficiently calculate the log-likelihood ratio (LLR). 
} to detect $M$-ary PSK FTN signaling is formally expressed as follows

\begin{enumerate}
	\item \textbf{Input:} The inputs to our SDR-based algorithm are the received samples ${\bm{y}}_{\textup{c}}$, the whitened matched filter impulse response, pulse shape $p(t)$ {(ISI matrix $\bm{G}$)}, and the Gaussian randomization number of iterations $L$. 
	\item Solve the relaxed convex optimization problem in in \eqref{eq:SDR_3} and find the optimal solutions ($\bm{A}^*$ and $\bm{a}^*$). 
	\item Generate random variable $\bm{\xi}_{\ell}$ as $\bm{\xi}_{\ell} \sim \mathcal{N}(\bm{a}^*,\bm{A}^* - \bm{a}^* \bm{a}^{*\textup{H}})$, $\ell = 1, ..., L$.
	\item Find $\ell_{\textup{op}}$ as in \eqref{eq:temp}.
	\item Calculate candidate suboptimal solutions to the FTN detection problem in \eqref{eq:SDP} as $\hat{\bm{a}} = \breve{\bm{a}}_{\ell_{\textup{op}}}$.
	\item \textbf{Output:} The estimated data symbol vector $\hat{\bm{a}}$.
\end{enumerate}

\subsection*{-- Complexity Analysis:} 
Solving the $M$-ary PSK FTN signaling detection problem in \eqref{eq:SDR_3} requires a complexity of $\mathcal{O}(N)^{3.5}$ \cite{luo2010semidefinite}; while the complexity of the Gaussian randomization step is $\mathcal{O}((N)^2 L)$ \cite{luo2010semidefinite}. Accordingly, the total computational complexity of the proposed algorithm is $\mathcal{O}((N)^{3.5} + (N)^2 L)$, which is polynomial in the received block length and is independent of the constellation size $M$ and the ISI length.


\section{Simulation Results} \label{sec:results}

The performance of our proposed algorithm to detect 8-ary PSK FTN signaling employing a root-raised cosine (rRC) pulse is investigated in this section. We assume a roll-off factor $\beta = 0.3$ and $0.5$, unless otherwise mentioned; and the number of iterations of Gaussian randomization $L$ is set to  1000.

\subsection{Performance of Our Proposed SDR-based Algorithm to Detect 8-ary FTN Signaling}

\begin{figure}[!t]
	\centering
	\includegraphics[width=0.50\textwidth]{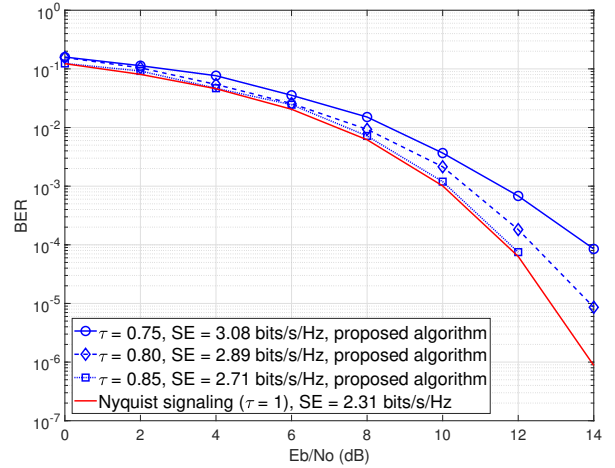}
	\caption{8-ary PSK FTN signaling performance of the proposed SDR-based algorithm for $\beta = 0.3$ and different values of  $\tau$.}\label{fig:Fig_1_beta_03_v2}
\end{figure}

The BER of 8-ary PSK FTN signaling of the proposed SDR-based scheme is shown in Fig. \ref{fig:Fig_1_beta_03_v2}. The performance is evaluated for roll-off factor $\beta = 0.3$ and for  spectral efficiency (SE) values of 3.08, 2.89, and 2.71 bits/s/Hz that correspond to $\tau$ values of 0.75, 0.80, and 0.85, respectively. One can notice that at SE of 2.71 bits/s/Hz ($\tau = 0.85$), our proposed SDR-based algorithm can completely remove the ISI and approach the BER of Nyquist transmission. This translates to an improvements of SE by 17.64$\%$ for the same roll-off factor value, SNR, and bandwidth. Similarly at BER = $10^{-3}$, it can be seen that improvements in SE by 33.33$\%$ and 25$\%$ are achieved for $\tau = 0.75$ and $0.8$ at a penalty of 0.6 and 1.4 dB in the SNR, respectively.

\begin{figure}[!t]
	\centering
	\includegraphics[width=0.50\textwidth]{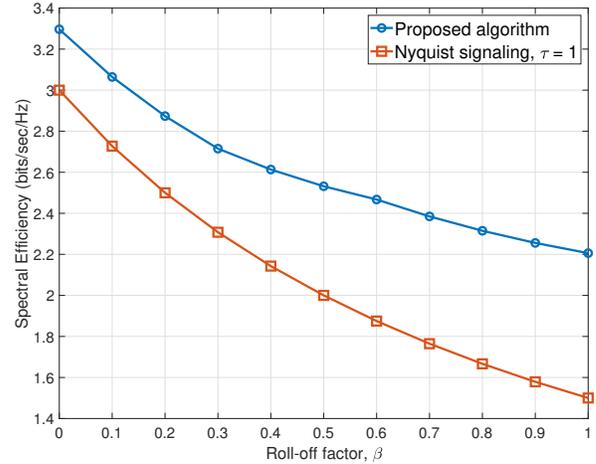}
	\caption{Effect of the roll-off factor of rRC pulse on the achieved SE by our proposed SDR-based algorithm at BER = $10^{-3}$.}\label{fig:Fig_3}
\end{figure}

Fig. \ref{fig:Fig_3} investigates the effect of the roll-off factor $\beta$ of the rRC pulse on the achieved SE of our proposed SDR-based algorithm at BER = $10^{-3}$. For a fair comparison with ISI-free transmission, we vary the value of $\tau$ (starting with a higher value of $\tau$ that is decremented) to maintain the same SNR achieved by the ISI-free transmission at  BER = $10^{-3}$.  First, one notices that the improvements of the SE achieved from our algorithm (when compared to its counterpart of ISI-free transmission) increases with increasing the value of $\beta$. This is in agreement to the conclusions reported in \cite{rusek2009constrained}. Additionally, the SE achieved by our proposed scheme for $\beta \in [0, 0.13]$ is higher than the maximum  SE of ISI-free transmission achieved at $\beta = 0$.

\subsection{Performance Comparison with the State-of-the-art Detectors}

\begin{figure}[!t]
	\centering
	\includegraphics[width=0.50\textwidth]{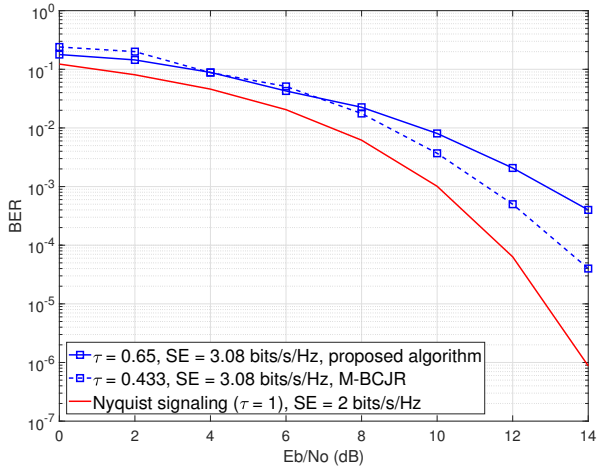}
	\caption{Comparison of our proposed SDR-based algorithm to detect 8-ary PSK FTN signaling and the M-BCJR algorithm in \cite{anderson2010turbo} to detect QPSK FTN signaling for $\beta = 0.5$ and the same SE.}\label{fig:Fig_2_beta_05_v4}
\end{figure}

We compare the performance of our proposed SDR-based algorithm to detect 8-ary FTN signaling to its counterpart of the M-BCJR algorithm in \cite{anderson2010turbo} to detect QPSK FTN signaling (i.e., both algorithms achieve the same SE of 3.08 bits/s/Hz for $\beta = 0.5$) in Fig. \ref{fig:Fig_2_beta_05_v4}. As can be seen, our proposed SDR-based algorithm (which is of polynomial complexity) is approximately at 1 dB SNR difference from the M-BCJR algorithm in \cite{anderson2010turbo} (which is of exponential complexity of the ISI length), at BER = $10^{-3}$. It is evident from Fig. \ref{fig:Fig_2_beta_05_v4} that our proposed SDR-based detection algorithm achieves a suboptimal solution that strikes a balance between complexity and performance. 

\section{Conclusion} \label{sec:conc}

FTN signaling showed merits in improving the spectral efficiency of future communication systems; however, the main challenge is the high complexity of its receivers.
In this paper, we have proposed a novel algorithm to estimate the transmit data symbols of any $M$-ary PSK FTN signaling. The proposed algorithm is based on concepts of semidefinite relaxation and Gaussian randomization and was shown to have polynomial complexity regardless of the ISI length or the constellation size $M$. Simulation results identified the favourable operating region of our proposed SDR-based scheme. For instance, improvements of the SE up to 33.33$\%$ can be achieved with a small increase of the SNR.
Additionally, it has been shown that the improvements of the SE increases with increasing the roll-off factor.


%
\IEEEpeerreviewmaketitle




%



\ifCLASSOPTIONcaptionsoff
  \newpage
\fi



%

\bibliographystyle{IEEEtran}
\bibliography{IEEEabrv,mybib_file}

%




\end{document}